\documentclass[twocolumn,journal]{article}
\usepackage[T1]{fontenc}
\usepackage[latin9]{inputenc}
\usepackage{booktabs}
\usepackage{calc}
\usepackage{mathtools}
\usepackage{amsmath}
\usepackage{amsthm}
\usepackage{amssymb}
\usepackage{graphicx}
\usepackage{wrapfig}
\usepackage[unicode=true,
 bookmarks=true,bookmarksnumbered=true,bookmarksopen=true,bookmarksopenlevel=1,
 breaklinks=false,pdfborder={0 0 0},pdfborderstyle={},backref=false,colorlinks=false]
 {hyperref}
\hypersetup{pdftitle={Your Title},
 pdfauthor={Your Name},
 pdfpagelayout=OneColumn, pdfnewwindow=true, pdfstartview=XYZ, plainpages=false}

\providecommand{\keywords}[1]
{
  \small	
  \textbf{\textit{Keywords---}} #1
}

\makeatletter

\providecommand{\tabularnewline}{\\}

\theoremstyle{plain}
\newtheorem{thm}{\protect\theoremname}
\theoremstyle{remark}
\newtheorem{rem}[thm]{\protect\remarkname}

\usepackage[caption=false,font=footnotesize]{subfig}

\makeatother

\providecommand{\remarkname}{Remark}
\providecommand{\theoremname}{Theorem}

\begin{document}
\title{Nonlinear $H_{\infty}$ Filtering on the Special Orthogonal Group
$SO(3)$ using Vector Directions}
\author{Farooq~Aslam and~M.~Farooq~Haydar\thanks{Farooq Aslam and Muhammad Farooq Haydar are with the Department of
Aeronautics and Astronautics, Institute of Space Technology, Islamabad,
Pakistan (e-mail: \protect\href{http://farooq.aslam87@gmail.com}{farooq.aslam87@gmail.com}
and \protect\href{http://mfarooqhaydar@gmail.com}{mfarooqhaydar@gmail.com})}}
\maketitle
\begin{abstract}
The problem of $H_{\infty}$ filtering for attitude estimation using
rotation matrices and vector measurements is studied. Starting from
a storage function on the Special Orthogonal Group $SO(3)$, a dissipation
inequality is considered, and a deterministic nonlinear $H_{\infty}$
filter is derived which respects a given upper bound $\gamma$ on
the energy gain from exogenous disturbances and initial estimation
errors to a generalized estimation error. The results are valid for
all estimation errors which correspond to an angular error of less
than $\pi/2$ radians in terms of the axis-angle representation. The
approach builds on earlier results on attitude estimation, in particular
nonlinear $H_{\infty}$ filtering using quaternions, and proposes
a novel filter developed directly on $SO(3)$. The proposed filter
employs the same innovation term as the Multiplicative Extended Kalman
Filter (MEKF), as well as a matrix gain updated in accordance with
a Riccati-type gain update equation. However, in contrast to the MEKF,
the proposed filter has an additional tuning gain, $\gamma$, which
enables it to be more aggressive during transients. The filter is
simulated for different conditions, and the results are compared with
those obtained using the continuous-time quaternion MEKF and the Geometric
Approximate Minimum Energy (GAME) filter. Simulations indicate competitive
performance. In particular, the GAME filter has the best transient
performance, followed by the proposed $H_{\infty}$ filter and the
quaternion MEKF. All three filters have similar steady-state performance.
Therefore, the proposed filter can be seen as a MEKF variant which
achieves better transient performance without significant degradation
in steady-state noise rejection.
\end{abstract}


\section{Introduction}

Rigid-body attitude estimation is a well-studied problem, and numerous solutions have been proposed \cite{markley2005nonlinear,crassidis2007survey}.
In general, attitude filtering problems are posed as optimization
problems which seek to minimize a suitably defined cost function indicative
of the filter's performance. The cost function usually comprises a
weighted average or integration of the estimation error, possibly
with a terminal cost. For linear systems, most optimal filtering solutions
are equivalent to Kalman filtering.

An alternative ($\textit{pessimistic}$, or perhaps $\textit{realistic}$)
approach to attitude filtering is given by worst-case or nonlinear
$H_{\infty}$ estimation. In formulating the $H_{\infty}$ filtering
problem, one optimizes not for the average estimation error, but for
the worst-case errors imparted by the worst-case disturbances and
large initial estimation errors. The resulting filter respects a given
upper bound on the energy gain from the disturbances and initial estimation
errors to a chosen performance measure, also a function of the estimation
error \cite{van1992sub,van1993nonlinear,isidori1994h,krener1994necessary,berman1996h,aguiar2009robust}.

Nonlinear $H_{\infty}$ filters have been applied successfully for
spacecraft attitude determination using quaternions \cite{crassidis2007survey,markley1993h,markley1994deterministic}.
Since $H_{\infty}$ filters minimize only the worst-case error and
not the mean square error (MSE), they are outperformed by the quaternion
Multiplicative Extended Kalman Filter (MEKF), particularly in problem
instances where the properties of the process and measurement noise/disturbance
are perfectly known. However, for the more practical case of uncertain
noise parameters leading to imperfect filter tuning, extensive Monte-Carlo
simulations show that the MEKF gradually loses its edge, and that
the $H_{\infty}$ filter performs better \cite{cooper2010spacecraft,choukroun2011spacecraft}.
Thus, nonlinear $H_{\infty}$ filtering offers a promising alternative
to optimal estimation methods. Nevertheless, the quaternion MEKF remains
the industry standard.

It is well-known that the attitude of a rigid body evolves on the
set of rotation matrices, the Special Orthogonal Group $SO(3)$, and
that all other attitude representations, including quaternions, suffer
either from a singularity or a sign ambiguity \cite{chaturvedi2011rigid}.
Due to the globally defined and unique attitude representation provided
by rotation matrices, the development of filtering solutions directly
on the Lie group $SO(3)$ has received considerable attention in recent
decades. The solutions proposed include nonlinear complementary filters
\cite{mahony2005complementary,mahony2008nonlinear,berkane2016deterministic,berkane2017design},
symmetry-preserving observers \cite{bonnabel2008symmetry,lageman2010gradient},
the Invariant Extended Kalman Filter (IEKF) \cite{bonnable2009invariant},
and minimum energy filters \cite{zamani2011near,zamani2012second,zamani2013minimum,ZamaniThesis,saccon2016second}.
Of these, the Geometric Approximate Minimum Energy (GAME) filter \cite{ZamaniThesis,zamani2013minimum},
is especially relevant since it is an optimal estimation method that
has been shown to significantly outperform the quaternion MEKF \cite{zamani2015nonlinear}.

Given the potential benefits of worst-case filtering and the increasing
interest in filtering algorithms on $SO(3)$, a promising solution
might be to combine nonlinear $H_{\infty}$ filtering with attitude
estimation using rotation matrices. This combination is the subject
of the present work. We note that there has been some preliminary
work in this direction. In particular, Lavoie et al. have proposed
an invariant extended $H_{\infty}$ filter for Lie groups \cite{lavoie2019invariant}.
Noting the similarities and crucial differences between the Extended
Kalman Filter (EKF) and its invariant counterpart, the IEKF, the authors
consider an analogous relationship between the Extended $H_{\infty}$
Filter and its potential invariant counterpart. This simple and insightful
observation leads them directly to an invariant extended $H_{\infty}$
filter for Lie groups. In contrast, we build on earlier results on
nonlinear worst-case filtering using quaternions \cite{markley1993h,markley1994deterministic},
and extend them to the case of attitude estimation using rotation
matrices. The main contribution of the paper is the development of
a novel nonlinear $H_{\infty}$ filter on the Lie group $SO(3)$.
It is envisaged that this exposition, which extends the preliminary
results of \cite{haydarCDC17}, will pave the way for future developments,
especially towards mixed $H_{2}/H_{\infty}$ filtering and possibly
even the generalized $H_{\infty}$ regulator problem on $SO(3)$ yielding
an estimator/controller pair which minimizes the impact of the worst-case
disturbances on the controlled variables.

The rest of the discussion is organized as follows. After Section
\ref{subsec:Notation} introduces the notation used in the paper,
Section \ref{sec:Problem-Formulation} poses the $H_{\infty}$ filtering
problem directly on $SO(3)$. Thereafter, the $H_{\infty}$ filter
is derived in Section \ref{sec:Filter-Derivation}, and its structure
is compared with that of the quaternion MEKF and the GAME filter.
In Section \ref{sec:Simulation-Results}, the proposed filter, the
MEKF, and the GAME filter are simulated for different test conditions.
Results indicate competitive performance. In particular, it is seen
that the GAME filter has the best transient performance, whereas the
$H_{\infty}$ filter demonstrates better convergence than the MEKF.
At steady-state, the three filters have similar performance. Finally,
in Section \ref{sec:Conclusion}, the main conclusions of the study
are reviewed.

\subsection{Notation\label{subsec:Notation}}

The Special Orthogonal group $SO(3)$ is the set of $3\times3$ orthogonal
matrices with determinant $1$, i.e.,
\begin{align*}
SO(3) & =\left\{ R\in\mathbb{R}^{3\times3}:R^{\top}R=I,\det\left(R\right)=1\right\} .
\end{align*}
The \textit{cross} operator $\times:\mathbb{R}^{3}\rightarrow\mathfrak{so}(3)$
transforms a vector in $\mathbb{R}^{3}$ to a $3$-by-$3$ skew-symmetric
matrix such that $x^{\times}y$ equals the cross product $x\times y$
for any $x,y\in\mathbb{R}^{3}$. In particular,
\begin{equation}
x^{\times}=\begin{bmatrix}0 & -x_{3} & x_{2}\\
x_{3} & 0 & -x_{1}\\
-x_{2} & x_{1} & 0
\end{bmatrix}.\label{eq:hatmap}
\end{equation}
The inverse of the cross operator is denoted by the \textit{vee} operator
$\vee:\mathfrak{so}(3)\rightarrow\mathbb{R}^{3}$. Lastly, we recall
the symmetric projection operator $\mathbb{P}_{s}:\mathbb{R}^{n\times n}\rightarrow\mathbb{R}^{n\times n}$
given by
\begin{equation}
\mathbb{P}_{s}(M)=\frac{M+M^{\top}}{2}.\label{eq:SymProj}
\end{equation}

\section{Problem Formulation\label{sec:Problem-Formulation}}

This section formulates the attitude estimation problem using rotation
matrices. In particular, we assume the following kinematic model:
\begin{equation}
\begin{split}\dot{R} & =R(\omega+g\delta)^{\times}\\
y_{i} & =R^{\top}r_{i}+k_{i}\varepsilon_{i},\quad i=1,\ldots,p,
\end{split}
\label{eq:System}
\end{equation}
where the state $R(t)$ evolves on $SO(3)$, and represents the orientation
of a body-fixed frame with respect to an inertial reference frame.
In the kinematics equation, the vectors $\omega,\delta\in\mathbb{R}^{3}$
are body-frame vectors denoting the measured angular velocity and
the process disturbance, respectively. In the output equation, the
vectors $\left\{ r_{i}\right\} $ specify known vector directions
in the reference frame, and the vectors $\left\{ y_{i}\right\} $
represent their measurements in the body-fixed frame with corresponding
measurement errors $\left\{ \varepsilon_{i}\right\} $. Lastly, the
coefficients $g$ and $\left\{ k_{i}\right\} $ scale the process
disturbance and measurement errors, respectively, and represent tuning
gains for the filter proposed in Section \ref{sec:Filter-Derivation}.

We consider the following estimator:
\begin{align}
\dot{\hat{R}} & =\hat{R}\left(\omega+G_{q}\tilde{y}\right)^{\times}, & \tilde{y}=\begin{bmatrix}y_{1}-\hat{y}_{1}\\
\vdots\\
y_{p}-\hat{y}_{p}
\end{bmatrix},\label{eq:Estimator_General}
\end{align}
where the filter gain $G_{q}\left(\hat{R},t\right)$ is a $3$-by-$3p$
matrix which will be specified later, and
\begin{equation}
\hat{y}_{i}=\hat{R}^{\top}r_{i},\label{eq:yi_hat}
\end{equation}
represents the estimate for the $i$-th reference direction $r_{i}$.
We note that this estimator structure is similar to the one considered
in \cite{markley1993h} for deriving a quaternion-based $H_{\infty}$
filter.

For the augmented system \eqref{eq:System}-\eqref{eq:Estimator_General},
the estimation error is defined as
\begin{equation}
\tilde{R}:=\hat{R}^{\top}R.\label{eq:R_tilde}
\end{equation}
Using the axis-angle representation, the estimation error can be expressed
as
\begin{align}
\tilde{R} & =I+\left(\sin\tilde{\theta}\right)e^{\times}+\left(1-\cos\tilde{\theta}\right)e^{\times}e^{\times},\label{eq:R_tilde_axang}
\end{align}
where $e\in\mathbb{S}^{2}$ denotes the axis of rotation between the
body-fixed and estimate frames, and $\tilde{\theta}$ denotes the
angle of rotation. We consider the following configuration error
vector as a penalty variable or $\textit{generalized}$ estimation
error:
\begin{align}
z & :=\frac{1}{2}\left(\tilde{R}-\tilde{R}^{\top}\right)^{\vee}.\label{eq:PenaltyVariable}
\end{align}
Using the axis-angle representation, the penalty variable can be expressed
as
\begin{equation}
z=e\sin\tilde{\theta}.\label{eq:PenaltyVariable_axang}
\end{equation}

We recall that the optimal $H_{\infty}$ filtering problem, for the
augmented system \eqref{eq:System}-\eqref{eq:Estimator_General},
seeks a filter gain which minimizes the $\mathcal{L}_{2}$ gain from
the process disturbance $\delta$, the measurement errors $\left\{ \varepsilon_{i}\right\} $,
and the initial conditions $R_{0}=R\left(0\right)$, $\hat{R}_{0}=\hat{R}\left(0\right)$
to the penalty variable $z\left(t\right)$. Since this is, in general,
a difficult problem to solve, we resort to the sub-optimal $H_{\infty}$
filtering problem which ensures that the $\mathcal{L}_{2}$ gain respects
a given upper bound $\gamma$, see e.g., \cite{van1993nonlinear,isidori1994h,krener1994necessary,markley1993h}.
More precisely, we seek a filter gain such that for all $t\geq0$,
$\delta$, $\left\{ \varepsilon_{i}\right\} $ , $R_{0}$, and $\hat{R}_{0}$,
the following dissipation inequality is satisfied:
\begin{align}
\int_{0}^{t}\left\Vert z\left(\tau\right)\right\Vert ^{2}d\tau & \leq\gamma^{2}\left\Vert z_{0}\right\Vert ^{2}+\gamma^{2}\int_{0}^{t}\Bigl[\left\Vert \delta\left(\tau\right)\right\Vert ^{2}\label{eq:suboptimal_Hinfty}\\
 & +\sum_{i}\left\Vert \varepsilon_{i}\right\Vert ^{2}\Bigr]d\tau.\nonumber
\end{align}
As has been well established in the literature on nonlinear $H_{\infty}$
control, see e.g., \cite{van1992sub,isidori1994h,krener1994necessary,markley1993h,markley1994deterministic},
the requirement \eqref{eq:suboptimal_Hinfty} on the $\mathcal{L}_{2}$
gain is closely related to the notion of dissipativity. In particular,
we seek a filter gain $G_{q}(\hat{R},t)$ and a smooth positive semidefinite
storage function $V(R,\hat{R},t)$ such that the following dissipation
inequality is satisfied:
\begin{equation}
\dot{V}\leq\gamma^{2}\left\Vert \delta\right\Vert ^{2}+\gamma^{2}\sum_{i}\left\Vert \varepsilon_{i}\right\Vert ^{2}-\left\Vert z\right\Vert ^{2}.\label{eq:DissipationInequality_A}
\end{equation}

Next, we obtain an expression for the time derivative of the penalty
variable $z\left(t\right)$. In particular, from \eqref{eq:PenaltyVariable},
we have that
\begin{equation}
z^{\times}=\frac{1}{2}\left(\tilde{R}-\tilde{R}^{\top}\right).\label{eq:z_cross}
\end{equation}
Conseqently, we can express the time derivative of the penalty variable
as
\begin{equation}
\left(\dot{z}\right)^{\times}=\frac{1}{2}\left(\dot{\tilde{R}}-\dot{\tilde{R}}^{\top}\right).\label{eq:zdot_A}
\end{equation}
Substituting the dynamics from \eqref{eq:System} and \eqref{eq:Estimator_General},
we can express the time derivative of $\tilde{R}$ as follows:
\begin{align*}
\dot{\tilde{R}} & =\hat{R}^{\top}\dot{R}+\dot{\hat{R}}^{\top}R,\\
 & =\hat{R}^{\top}R\left(\omega+g\delta\right)^{\times}-\left(\omega+G_{q}\tilde{y}\right)^{\times}\hat{R}^{\top}R,\\
\implies\dot{\tilde{R}} & =\tilde{R}\omega^{\times}-\omega^{\times}\tilde{R}+g\tilde{R}\delta^{\times}-\left(G_{q}\tilde{y}\right)^{\times}\tilde{R}.
\end{align*}
Consequently, we have that:
\begin{align*}
\dot{\tilde{R}}-\dot{\tilde{R}}^{\top} & =\left(\tilde{R}\omega^{\times}+\omega^{\times}\tilde{R}^{\top}\right)-\left(\omega^{\times}\tilde{R}+\tilde{R}^{\top}\omega^{\times}\right)\\
 & \quad+g\left(\tilde{R}\delta^{\times}+\delta^{\times}\tilde{R}^{\top}\right)-\left(\left(G_{q}\tilde{y}\right)^{\times}\tilde{R}+\tilde{R}^{\top}\left(G_{q}\tilde{y}\right)^{\times}\right).
\end{align*}
We apply the following identity to each term on the right-hand side
of the above expression:
\[
x^{\times}A+A^{\top}x^{\times}=\left(\left\{ \text{tr}[A]I-A\right\} x\right)^{\times},\quad x\in\mathbb{R}^{3},A\in\mathbb{R}^{3\times3}.
\]
For the first two terms, we observe that:
\begin{align*}
 & \left(\tilde{R}\omega^{\times}+\omega^{\times}\tilde{R}^{\top}\right)-\left(\omega^{\times}\tilde{R}+\tilde{R}^{\top}\omega^{\times}\right)\\
= & \left(\left(\text{tr}\left[\tilde{R}^{\top}\right]I-\tilde{R}^{\top}-\text{tr}\left[\tilde{R}\right]I+\tilde{R}\right)\omega\right)^{\times},\\
= & \left(\left(\tilde{R}-\tilde{R}^{\top}\right)\omega\right)^{\times},\\
= & 2\left(z^{\times}\omega\right)^{\times},
\end{align*}
In the above simplification, the last equality follows from the relation
\eqref{eq:z_cross}. Next, we introduce the following mapping:
\begin{equation}
\textbf{E}\left(M\right):=\text{tr}\left[M\right]I-M^{\top}.\label{eq:E}
\end{equation}
Consequently, we can express \eqref{eq:zdot_A} as:
\[
\left(\dot{z}\right)^{\times}=\left(z^{\times}\omega\right)^{\times}+\frac{g}{2}\left(\textbf{E}\left(\tilde{R}\right)\delta\right)^{\times}-\frac{1}{2}\left(\textbf{E}^{\top}\left(\tilde{R}\right)G_{q}\tilde{y}\right)^{\times}.
\]
Removing the $\times$ operator, and substituting for $\tilde{y}$
and $\hat{y}_{i}$ from \eqref{eq:Estimator_General}-\eqref{eq:yi_hat},
we obtain:
\begin{align}
\dot{z} & =z^{\times}\omega+\frac{g}{2}\textbf{E}\left(\tilde{R}\right)\delta-\frac{1}{2}\textbf{E}^{\top}\left(\tilde{R}\right)G_{q}\begin{bmatrix}R^{\top}r_{1}-\hat{R}^{\top}r_{1}\\
\vdots\\
R^{\top}r_{p}-\hat{R}^{\top}r_{p}
\end{bmatrix}\nonumber \\
 & \quad-\frac{1}{2}\textbf{E}^{\top}\left(\tilde{R}\right)G_{q}\begin{bmatrix}k_{1}\varepsilon_{1}\\
\vdots\\
k_{p}\varepsilon_{p}
\end{bmatrix}.\label{eq:zdot_B}
\end{align}

\section{Filter Derivation\label{sec:Filter-Derivation}}

This section presents a novel $H_{\infty}$ filter for attitude estimation
using rotation matrices. The filter derivation follows the same procedure
as is used in \cite{markley1993h,markley1994deterministic} for deriving
a quaternion-based $H_{\infty}$ filter. In particular, starting from
a storage function on $SO\left(3\right)$, a dissipation inequality
of the form \eqref{eq:DissipationInequality_A} is considered. The
worst-case process disturbance and measurement errors are obtained,
as well as the gain which optimizes for the worst-case disturbances.
These values are used to simplify the dissipation inequality. Thereafter,
for sufficiently small estimation errors, a deterministic nonlinear
$H_{\infty}$ filter is derived, the filter gain $G_{q}$ is specified,
and a Riccati-type gain update equation is obtained.

We consider the following storage function:
\begin{equation}
V(R,\hat{R},t)=\gamma^{2}z^{\top}P^{-1}(t)z,\label{eq:StorageFunction}
\end{equation}
where the penalty variable $z(t)$ is defined as in \eqref{eq:PenaltyVariable},
and $P(t)$ is a symmetric, positive definite, $3$-by-$3$ matrix.
The time derivative of the storage function is given by
\begin{equation}
\dot{V}=-\gamma^{2}z^{\top}P^{-1}\dot{P}P^{-1}z+2\gamma^{2}z^{\top}P^{-1}\dot{z}.\label{eq:Vdot}
\end{equation}
Substituting \eqref{eq:zdot_B} in \eqref{eq:Vdot}, the dissipation
inequality \eqref{eq:DissipationInequality_A} can be expressed as:
\begin{align}
 & -\gamma^{2}z^{\top}P^{-1}\dot{P}P^{-1}z-2\gamma^{2}z^{\top}P^{-1}\omega^{\times}z\nonumber \\
 & +g\gamma^{2}z^{\top}P^{-1}\textbf{E}\left(\tilde{R}\right)\delta\nonumber \\
 & -\gamma^{2}z^{\top}P^{-1}\textbf{E}^{\top}\left(\tilde{R}\right)G_{q}\begin{bmatrix}R^{\top}r_{1}-\hat{R}^{\top}r_{1}\\
\vdots\\
R^{\top}r_{p}-\hat{R}^{\top}r_{p}
\end{bmatrix}\nonumber \\
 & -\gamma^{2}z^{\top}P^{-1}\textbf{E}^{\top}\left(\tilde{R}\right)G_{q}\begin{bmatrix}k_{1}\varepsilon_{1}\\
\vdots\\
k_{p}\varepsilon_{p}
\end{bmatrix}\nonumber \\
 & -\gamma^{2}\left\Vert \delta\right\Vert ^{2}-\gamma^{2}\sum_{i}\left\Vert \varepsilon_{i}\right\Vert ^{2}+z^{\top}z\leq0.\label{eq:DissipationInequality_B}
\end{align}

\begin{rem}
Using the axis-angle representation \eqref{eq:PenaltyVariable_axang}
of the penalty variable, the storage function \eqref{eq:StorageFunction}
can be expressed as
\begin{equation}
V=\gamma^{2}\left(\sin^{2}\tilde{\theta}\right)e^{\top}P^{-1}e.\label{eq:StorageFunction_Axang}
\end{equation}
Thus, we have that the angular term increases only in the interval
$|\tilde{\theta}|\in\left(0,\frac{\pi}{2}\right)$ and decreases to
$0$ thereafter. This means that although the storage function is
nonnegative, it can be used to obtain stability and performance guarantees
only for angular errors up to $\pi/2$ radians. Having noted this
limitation, we now proceed to the main steps in the filter derivation.
\end{rem}

\subsection{Worst-case Disturbances}

The dissipation inequality \eqref{eq:DissipationInequality_B} is
quadratic in the process disturbance $\delta$ and the measurement
errors $\left\{ \varepsilon_{i}\right\} $. In particular, the worst-case
process disturbance can be obtained as follows:
\begin{align}
 & g\gamma^{2}\textbf{E}^{\top}\left(\tilde{R}\right)P^{-1}z-2\gamma^{2}\delta^{*}=0,\nonumber \\
 & \implies\delta^{*}=\frac{g}{2}\textbf{E}^{\top}\left(\tilde{R}\right)P^{-1}z.\label{eq:WorstCase_ProcessDisturbance}
\end{align}
Its contribution to the dissipation inequality \eqref{eq:DissipationInequality_B}
is given by the term
\[
\frac{\gamma^{2}g^{2}}{4}z^{\top}P^{-1}\textbf{E}\left(\tilde{R}\right)\textbf{E}^{\top}\left(\tilde{R}\right)P^{-1}z.
\]

Next, we partition the filter gain $G_{q}$ as
\[
G_{q}=\begin{bmatrix}G_{q,1} & \cdots & G_{q,p}\end{bmatrix},
\]
where $G_{q,i}\in\mathbb{R}^{3\times3}$. Then, the $i$-th worst-case
measurement error can be obtained as:
\begin{align}
 & -k_{i}\gamma^{2}G_{q,i}^{\top}\textbf{E}\left(\tilde{R}\right)P^{-1}z-2\gamma^{2}\varepsilon_{i}^{*}=0,\nonumber \\
 & \implies\varepsilon_{i}^{*}=-\frac{k_{i}}{2}G_{q,i}^{\top}\textbf{E}\left(\tilde{R}\right)P^{-1}z.\label{eq:WorstCase_MeasurementError}
\end{align}
Its contribution to \eqref{eq:DissipationInequality_B} is given by
\[
\frac{\gamma^{2}k_{i}^{2}}{4}z^{\top}P^{-1}\textbf{E}^{\top}\left(\tilde{R}\right)G_{q,i}G_{q,i}^{\top}\textbf{E}\left(\tilde{R}\right)P^{-1}z,
\]
and the total contribution of the worst-case measurement errors $\left\{ \varepsilon_{i}^{*}\right\} $
is given by
\[
\frac{\gamma^{2}}{4}z^{\top}P^{-1}\textbf{E}^{\top}\left(\tilde{R}\right)G_{q}KG_{q}^{\top}\textbf{E}\left(\tilde{R}\right)P^{-1}z,
\]
where
\begin{equation}
K:=\text{diag}\left[k_{1}^{2}I,\ldots,k_{p}^{2}I\right]\in\mathbb{R}^{3p\times3p}.\label{eq:K}
\end{equation}
Consequently, after substituting the worst-case process disturbance
$\delta^{*}$ and the worst-case measurement errors $\left\{ \varepsilon_{i}^{*}\right\} $,
we arrive at the following dissipation inequality:
\begin{align}
 & -\gamma^{2}z^{\top}P^{-1}\dot{P}P^{-1}z-2\gamma^{2}z^{\top}P^{-1}\omega^{\times}z\nonumber \\
 & +\frac{\gamma^{2}g^{2}}{4}z^{\top}P^{-1}\textbf{E}\left(\tilde{R}\right)\textbf{E}^{\top}\left(\tilde{R}\right)P^{-1}z\nonumber \\
 & -\gamma^{2}z^{\top}P^{-1}\textbf{E}^{\top}\left(\tilde{R}\right)G_{q}\begin{bmatrix}R^{\top}r_{1}-\hat{R}^{\top}r_{1}\\
\vdots\\
R^{\top}r_{p}-\hat{R}^{\top}r_{p}
\end{bmatrix}\nonumber \\
 & +\frac{\gamma^{2}}{4}z^{\top}P^{-1}\textbf{E}^{\top}\left(\tilde{R}\right)G_{q}KG_{q}^{\top}\textbf{E}\left(\tilde{R}\right)P^{-1}z\nonumber \\
 & +z^{\top}z\leq0.\label{eq:DissipationInequality_C}
\end{align}

\subsection{Filter Gain}

In order to find the optimal filter gain for the worst-case disturbances,
we perform completion of squares for the terms in the dissipation
inequality \eqref{eq:DissipationInequality_C} that contain the gain
$G_{q}$. In particular, these terms can be re-stated as:
\begin{align*}
 & \frac{\gamma^{2}}{4}\left\Vert K^{1/2}G_{q}^{\top}\textbf{E}\left(\tilde{R}\right)P^{-1}z-2K^{-1/2}\begin{bmatrix}R^{\top}r_{1}-\hat{R}^{\top}r_{1}\\
\vdots\\
R^{\top}r_{p}-\hat{R}^{\top}r_{p}
\end{bmatrix}\right\Vert ^{2}\\
 & -\gamma^{2}\begin{bmatrix}R^{\top}r_{1}-\hat{R}^{\top}r_{1}\\
\vdots\\
R^{\top}r_{p}-\hat{R}^{\top}r_{p}
\end{bmatrix}^{\top}K^{-1}\begin{bmatrix}R^{\top}r_{1}-\hat{R}^{\top}r_{1}\\
\vdots\\
R^{\top}r_{p}-\hat{R}^{\top}r_{p}
\end{bmatrix}.
\end{align*}
Therefore, the gain $G_{q}$ which minimizes the left-hand side of
\eqref{eq:DissipationInequality_C} satisfies the following condition:
\begin{equation}
z^{\top}P^{-1}\textbf{E}^{\top}\left(\tilde{R}\right)G_{q}=2\begin{bmatrix}R^{\top}r_{1}-\hat{R}^{\top}r_{1}\\
\vdots\\
R^{\top}r_{p}-\hat{R}^{\top}r_{p}
\end{bmatrix}^{\top}K^{-1}.\label{eq:min_Gq}
\end{equation}
Substituting $\hat{y}_{i}$ and $\tilde{R}$ from \eqref{eq:yi_hat}-\eqref{eq:R_tilde},
we observe that:
\[
R^{\top}r_{i}-\hat{R}^{\top}r_{i}=\left(R^{\top}\hat{R}-I\right)\hat{R}^{\top}r_{i}=\left(\tilde{R}^{\top}-I\right)\hat{y}_{i}.
\]
Furthermore, using the axis-angle representations \eqref{eq:R_tilde_axang}
and \eqref{eq:PenaltyVariable_axang}, and assuming that $\cos\tilde{\theta}\neq-1$,
we proceed as follows:
\begin{align}
\left(\tilde{R}^{\top}-I\right)\hat{y}_{i} & =\left[\left(1-\cos\tilde{\theta}\right)e^{\times}e^{\times}-\left(\sin\tilde{\theta}\right)e^{\times}\right]\hat{y}_{i},\nonumber \\
 & =\left[\left(\frac{\sin\tilde{\theta}}{1+\cos\tilde{\theta}}\right)e^{\times}-I\right]\left(\sin\tilde{\theta}\right)e^{\times}\hat{y}_{i},\nonumber \\
 & =\left(\frac{1}{1+\cos\tilde{\theta}}z^{\times}-I\right)z^{\times}\hat{y}_{i},\nonumber \\
 & =-\left(\frac{1}{1+\cos\tilde{\theta}}z^{\times}-I\right)\left(\hat{y}_{i}\right)^{\times}z.\label{eq:Temp}
\end{align}
Therefore, using the preceding relations, the minimizing gain condition
can be expressed as follows:
\begin{align*}
z^{\top}P^{-1}\textbf{E}^{\top}\left(\tilde{R}\right)G_{q} & =-2z^{\top}\left[\begin{array}{cc}
\left(\hat{y}_{1}\right)^{\times}\left(\frac{1}{1+\cos\tilde{\theta}}z^{\times}+I\right) & \cdots\end{array}\right.\\
 & \left.\begin{array}{cc}
\ldots & \left(\hat{y}_{p}\right)^{\times}\left(\frac{1}{1+\cos\tilde{\theta}}z^{\times}+I\right)\end{array}\right]K^{-1}.
\end{align*}
This leads to the following sufficient condition for the minimizing
gain:
\begin{align*}
\textbf{E}^{\top}\left(\tilde{R}\right)G_{q} & =-2P\left(t\right)\left[\begin{array}{cc}
\left(\hat{y}_{1}\right)^{\times}\left(\frac{1}{1+\cos\tilde{\theta}}z^{\times}+I\right) & \cdots\end{array}\right.\\
 & \left.\begin{array}{cc}
\ldots & \left(\hat{y}_{p}\right)^{\times}\left(\frac{1}{1+\cos\tilde{\theta}}z^{\times}+I\right)\end{array}\right]K^{-1}.
\end{align*}
We note that this condition couples the gain matrix $G_{q}$ to the
true estimation error $\tilde{R}=\hat{R}^{\top}R$, and that the latter
cannot be computed since the true trajectory $R(t)$ is not available.
However, an approximate solution may be obtained for small estimation
errors, i.e., for $\hat{R}(t)\approx R(t)$. In particular, as $\tilde{R}(t)\rightarrow I$,
we have that the penalty variable $z\rightarrow0$, the mapping $\textbf{E}\left(\tilde{R}\right)\rightarrow2I$,
and the sufficient condition for the minimizing gain simplifies thus:
\begin{align}
G_{q}\left(\hat{R},t\right) & =-P\left(t\right)\begin{bmatrix}\left(\hat{R}^{\top}r_{1}\right)^{\times} & \ldots & \left(\hat{R}^{\top}r_{p}\right)^{\times}\end{bmatrix}K^{-1}\nonumber \\
 & =P\left(t\right)H^{\top}\left(\hat{R},r\right)K^{-1},\label{eq:Gq}
\end{align}
where
\begin{equation}
H\left(\hat{R},r\right):=\begin{bmatrix}\left(\hat{R}^{\top}r_{1}\right)^{\times}\\
\vdots\\
\left(\hat{R}^{\top}r_{p}\right)^{\times}
\end{bmatrix}.\label{eq:H}
\end{equation}
Consequently, substituting $K$ and $G_{q}$ from \eqref{eq:K} and
\eqref{eq:Gq}, we obtain the following expression for the estimator
\eqref{eq:Estimator_General}:
\begin{equation}
\dot{\hat{R}}=\hat{R}\left(\omega-P\left(t\right)\sum_{i}k_{i}^{-2}\left(\hat{R}^{\top}r_{i}\right)\times y_{i}\right)^{\times}.\label{eq:Estimator_Particular}
\end{equation}
We note that the same estimator structure is used in \cite{zamani2012second},
and admits, as special cases, the structures employed in \cite{mahony2008nonlinear,berkane2017design}
for attitude estimation using vector measurements.

\subsection{Gain Update Equation}

This section derives a Riccati-type gain update equation for the time-varying
filter gain $P\left(t\right)$ in the estimator \eqref{eq:Estimator_Particular}.
Substituting the minimizing gain condition \eqref{eq:min_Gq} in the
dissipation inequality \eqref{eq:DissipationInequality_C} leads to
the following expression:
\begin{align}
 & -\gamma^{2}z^{\top}P^{-1}\dot{P}P^{-1}z-2\gamma^{2}z^{\top}P^{-1}\omega^{\times}z\nonumber \\
 & -\gamma^{2}\begin{bmatrix}R^{\top}r_{1}-\hat{R}^{\top}r_{1}\\
\vdots\\
R^{\top}r_{p}-\hat{R}^{\top}r_{p}
\end{bmatrix}^{\top}K^{-1}\begin{bmatrix}R^{\top}r_{1}-\hat{R}^{\top}r_{1}\\
\vdots\\
R^{\top}r_{p}-\hat{R}^{\top}r_{p}
\end{bmatrix}\nonumber \\
 & +\frac{\gamma^{2}g^{2}}{4}z^{\top}P^{-1}\textbf{E}\left(\tilde{R}\right)\textbf{E}^{\top}\left(\tilde{R}\right)P^{-1}z\nonumber \\
 & +z^{\top}z\leq0\label{eq:HJI}
\end{align}
The next step is to re-state each term in the above dissipation inequality
using the axis-angle representation $\left(e,\tilde{\theta}\right)$
for the estimation error $\tilde{R}$. For the second term , we observe
that:
\begin{align}
-2\gamma^{2}z^{\top}P^{-1}\omega^{\times}z & =-\gamma^{2}z^{\top}P^{-1}\omega^{\times}z+\gamma^{2}z^{\top}\omega^{\times}P^{-1}z\nonumber \\
 & =\gamma^{2}z^{\top}\left(\omega^{\times}P^{-1}-P^{-1}\omega^{\times}\right)z\nonumber \\
 & =\gamma^{2}\left(\sin^{2}\tilde{\theta}\right)e^{\top}\left(\omega^{\times}P^{-1}-P^{-1}\omega^{\times}\right)e\label{eq:Second}
\end{align}

In order to simplify the third term in \eqref{eq:HJI}, we recall
the relation \eqref{eq:Temp} and the following identities:
\begin{align*}
z^{\times}z^{\times} & =zz^{\top}-\left(\sin^{2}\tilde{\theta}\right)I, & z^{\top}z^{\times}=0.
\end{align*}
Assuming that $\cos\tilde{\theta}\neq-1$, we proceed as follows:
\begin{align*}
 & \left(R^{\top}r_{i}-\hat{R}^{\top}r_{i}\right)^{\top}\left(R^{\top}r_{i}-\hat{R}^{\top}r_{i}\right)\\
 & =z^{\top}\left(\hat{R}^{\top}r_{i}\right)^{\times}\left(\frac{1}{\left(1+\cos\tilde{\theta}\right)^{2}}z^{\times}z^{\times}-I\right)\left(\hat{R}^{\top}r_{i}\right)^{\times}z\\
 & =\left(-\frac{\sin^{2}\tilde{\theta}}{\left(1+\cos\tilde{\theta}\right)^{2}}-1\right)z^{\top}\left(\hat{R}^{\top}r_{i}\right)^{\times}\left(\hat{R}^{\top}r_{i}\right)^{\times}z\\
 & =-\frac{2}{1+\cos\tilde{\theta}}z^{\top}\left(\hat{R}^{\top}r_{i}\right)^{\times}\left(\hat{R}^{\top}r_{i}\right)^{\times}z\\
 & =-2\left(1-\cos\tilde{\theta}\right)e^{\top}\left(\hat{R}^{\top}r_{i}\right)^{\times}\left(\hat{R}^{\top}r_{i}\right)^{\times}e
\end{align*}
Consequently, using the axis-angle representation and the definition
of $H$ in \eqref{eq:H}, the third term in \eqref{eq:HJI} can be
simplified as follows:
\begin{align}
 & -\gamma^{2}\begin{bmatrix}R^{\top}r_{1}-\hat{R}^{\top}r_{1}\\
\vdots\\
R^{\top}r_{p}-\hat{R}^{\top}r_{p}
\end{bmatrix}^{\top}K^{-1}\begin{bmatrix}R^{\top}r_{1}-\hat{R}^{\top}r_{1}\\
\vdots\\
R^{\top}r_{p}-\hat{R}^{\top}r_{p}
\end{bmatrix}\nonumber \\
 & =2\gamma^{2}\left(1-\cos\tilde{\theta}\right)e^{\top}\left(\sum_{i}k_{i}^{-2}\left(\hat{R}^{\top}r_{i}\right)^{\times}\left(\hat{R}^{\top}r_{i}\right)^{\times}\right)e\nonumber \\
 & =-2\gamma^{2}\left(1-\cos\tilde{\theta}\right)e^{\top}H^{\top}K^{-1}He\label{eq:Third}
\end{align}
For the fourth term in \eqref{eq:HJI}, we recall the map \eqref{eq:E}
and proceed as follows:
\begin{align*}
\textbf{E}\left(\tilde{R}\right)\textbf{E}^{\top}\left(\tilde{R}\right) & =\text{tr}^{2}\left[\tilde{R}\right]I-\text{tr}\left[\tilde{R}\right]\left(\tilde{R}+\tilde{R}^{\top}\right)+I\\
 & =I+\text{tr}\left[\tilde{R}\right]\left[\text{tr}\left[\tilde{R}\right]I-\left(\tilde{R}+\tilde{R}^{\top}\right)\right]
\end{align*}
Using the axis-angle representation \eqref{eq:R_tilde_axang}, we
observe that:
\begin{align*}
 & \text{tr}\left[\tilde{R}\right]I-\left(\tilde{R}+\tilde{R}^{\top}\right)\\
 & =\left(1+2\cos\tilde{\theta}\right)I-2\left[I+\left(1-\cos\tilde{\theta}\right)e^{\times}e^{\times}\right]\\
 & =\left(2\cos\tilde{\theta}-1\right)I-2\left(1-\cos\tilde{\theta}\right)e^{\times}e^{\times}\\
 & =\left(2\cos\tilde{\theta}-1\right)I-2\left(1-\cos\tilde{\theta}\right)\left(ee^{\top}-I\right)\\
 & =I-2\left(1-\cos\tilde{\theta}\right)ee^{\top}
\end{align*}
Therefore:
\begin{align*}
\textbf{E}\left(\tilde{R}\right)\textbf{E}^{\top}\left(\tilde{R}\right) & =2\left(1+\cos\tilde{\theta}\right)I\\
 & \quad-2\left(1+2\cos\tilde{\theta}\right)\left(1-\cos\tilde{\theta}\right)ee^{\top}
\end{align*}
Consequently, the fourth term in \eqref{eq:HJI} can be expressed
as follows:
\begin{align}
 & \frac{\gamma^{2}g^{2}}{2}\left[\left(1+\cos\tilde{\theta}\right)\left(\sin^{2}\tilde{\theta}\right)e^{\top}P^{-2}e\right.\nonumber \\
 & \left.-\left(1+2\cos\tilde{\theta}\right)\left(1-\cos\tilde{\theta}\right)\left(\sin^{2}\tilde{\theta}\right)\left(e^{\top}P^{-1}e\right)^{2}\right]\label{eq:Fourth}
\end{align}
Finally, using \eqref{eq:Second}-\eqref{eq:Fourth}, we re-state
the dissipation inequality \eqref{eq:HJI} in terms of the axis-angle
representation as follows:
\begin{align}
 & -\gamma^{2}\left(\sin^{2}\tilde{\theta}\right)e^{\top}P^{-1}\dot{P}P^{-1}e\nonumber \\
 & +\gamma^{2}\left(\sin^{2}\tilde{\theta}\right)e^{\top}\left(\omega^{\times}P^{-1}-P^{-1}\omega^{\times}\right)e\nonumber \\
 & -2\gamma^{2}\left(1-\cos\tilde{\theta}\right)e^{\top}H^{\top}K^{-1}He\nonumber \\
 & +\frac{\gamma^{2}g^{2}}{2}\left(1+\cos\tilde{\theta}\right)\left(\sin^{2}\tilde{\theta}\right)e^{\top}P^{-2}e+\left(\sin^{2}\tilde{\theta}\right)e^{\top}e\nonumber \\
 & -\frac{\gamma^{2}g^{2}}{2}\left(1+2\cos\tilde{\theta}\right)\left(1-\cos\tilde{\theta}\right)\left(\sin^{2}\tilde{\theta}\right)\left(e^{\top}P^{-1}e\right)^{2}\leq0\label{eq:HJI_Axang}
\end{align}

In the above dissipation inequality, the first five terms are second-order
in the Euler axis $e$, the last term is fourth-order, and the coefficient
of the last term is positive for $|\tilde{\theta}|\leq120^{\circ}$.
Furthermore, we observe that:
\[
\frac{1}{2}\sin^{2}\tilde{\theta}\leq1-\cos\tilde{\theta}
\]
Therefore, for $|\tilde{\theta}|\leq90^{\circ}$, a sufficient condition
for the dissipation inequality \eqref{eq:HJI_Axang} to hold is given
as follows:
\begin{align*}
 & -\gamma^{2}\left(\sin^{2}\tilde{\theta}\right)e^{\top}P^{-1}\dot{P}P^{-1}e\\
 & +\gamma^{2}\left(\sin^{2}\tilde{\theta}\right)e^{\top}\left(\omega^{\times}P^{-1}-P^{-1}\omega^{\times}\right)e\\
 & -\gamma^{2}\left(\sin^{2}\tilde{\theta}\right)e^{\top}H^{\top}K^{-1}He\\
 & +\gamma^{2}g^{2}\left(\sin^{2}\tilde{\theta}\right)e^{\top}P^{-2}e+\left(\sin^{2}\tilde{\theta}\right)e^{\top}e\leq0
\end{align*}
Dividing throughout by $\gamma^{2}\left(\sin^{2}\tilde{\theta}\right)$
and considering only the matrix-valued terms, we obtain the following
Riccati-type update equation for the filter gain $P\left(t\right)$:
\begin{align*}
\dot{P} & =P\omega^{\times}-\omega^{\times}P-PH^{\top}K^{-1}HP+g^{2}I+\frac{1}{\gamma^{2}}P^{2}
\end{align*}
Substituting for $\hat{y}_{i}$ and $H$ from \eqref{eq:yi_hat}
and \eqref{eq:H}, respectively, we summarize the equations for the
continuous-time $H_{\infty}$ filter on $SO\left(3\right)$ as follows:
\begin{equation}
\begin{split}\dot{\hat{R}} & =\hat{R}\left(\omega-Pl\right)^{\times},\quad l=\sum_{i}k_{i}^{-2}\left(\hat{y}_{i}\times y_{i}\right)\\
\dot{P} & =\mathbb{P}_{s}\left(2P\omega^{\times}\right)+P\left(\sum_{i}k_{i}^{-2}\hat{y}_{i}^{\times}\hat{y}_{i}^{\times}\right)P+g^{2}I+\frac{1}{\gamma^{2}}P^{2},
\end{split}
\label{eq:EHF}
\end{equation}
where $\mathbb{P}_{s}$ denotes the symmetric projection operator
\eqref{eq:SymProj}. We note that the gain update equation is similar
to that proposed for the quaternion $H_{\infty}$ filter \cite{markley1993h,markley1994deterministic}.
\begin{rem}
The proposed $H_{\infty}$ filter \eqref{eq:EHF} is similar to the
quaternion MEKF \cite{markley2004multiplicative} and the quaternion
left-invariant EKF (LIEKF) \cite{gui2018quaternion}. More precisely,
suppose that in the kinematic system under consideration \eqref{eq:System},
the process disturbance $\delta$ and the measurement errors $\left\{ \varepsilon_{i}\right\} $
are zero-mean Gaussian noises with standard deviations $\sigma_{\omega}$
and $\left\{ \sigma_{i}\right\} $, respectively. Then, the continuous-time
quaternion MEKF \cite{markley2004multiplicative}, which also coincides
with the continuous-time quaternion LIEKF \cite{gui2018quaternion},
can be expressed as:
\begin{equation}
\begin{split}\frac{d}{dt}\begin{bmatrix}\hat{q}_{v}\\
\hat{q}_{s}
\end{bmatrix} & =\frac{1}{2}\begin{bmatrix}\hat{q}_{s}I+\hat{q}_{v}^{\times}\\
-\hat{q}_{v}^{\top}
\end{bmatrix}\omega_{\text{ref}},\\
\omega_{\text{ref}} & =\omega-P\sum_{i}\sigma_{i}^{-2}\left(\hat{y}_{i}\times y_{i}\right),
\end{split}
\label{eq:MEKF_Estimator}
\end{equation}
where
\begin{equation}
\hat{R}=\left(\hat{q}_{s}^{2}-\hat{q}_{v}^{\top}\hat{q}_{v}\right)I+2\hat{q}_{v}\hat{q}_{v}^{\top}+2\hat{q}_{s}\hat{q}_{v}^{\times},\label{eq:MEKF_Rhat}
\end{equation}
and the filter gain is updated as follows:
\begin{equation}
\dot{P}=\mathbb{P}_{s}\left(2P\omega^{\times}\right)+P\left(\sum_{i}\sigma_{i}^{-2}\hat{y}_{i}^{\times}\hat{y}_{i}^{\times}\right)P+\sigma_{\omega}^{2}I.\label{eq:Riccati_MEKF}
\end{equation}
Comparing \eqref{eq:EHF} and \eqref{eq:MEKF_Estimator}-\eqref{eq:Riccati_MEKF},
we note that the proposed $H_{\infty}$ filter and the continuous-time
quaternion MEKF/LIEKF use identical innovation terms. Likewise, the
gain update equations are similar except that the proposed $H_{\infty}$
filter has an additional second-order term containing $\gamma$. This
term can be seen as an extra tuning gain which can be used to make
the filter more aggressive during transients. Therefore, the proposed
filter can be seen as a MEKF variant which achieves better transient
performance without significant degradation in steady-state noise
rejection. Moreover, this term vanishes in the limit $\gamma\rightarrow\infty$.
\end{rem}
\begin{rem}
The vector directions $H_{\infty}$ filter \eqref{eq:EHF} can also
be compared with the Geometric Approximate Minimum Energy (GAME) filter
\cite{zamani2012second,ZamaniThesis}. The GAME filter is especially
relevant since it is an optimal estimation method that has been shown
to significantly outperform the quaternion MEKF \cite{zamani2015nonlinear}.
For the kinematic system under consideration \eqref{eq:System}, the
GAME filter \cite[Section 4.3]{ZamaniThesis}, \cite{izadi2015comparison}
can be expressed as:
\begin{equation}
\begin{split}\dot{\hat{R}} & =\hat{R}\left(\omega-Pl\right)^{\times},\quad l=\sum_{i}k_{i}^{-2}\left(\hat{y}_{i}\times y_{i}\right),\\
\dot{P} & =\mathbb{P}_{s}\left(2P\omega^{\times}\right)+P\left(\sum_{i}k_{i}^{-2}\hat{y}_{i}^{\times}\hat{y}_{i}^{\times}\right)P+g^{2}I\\
 & \quad-\mathbb{P}_{s}\left(P\left(Pl\right)^{\times}\right)+P\;\textbf{E}\left(\sum_{i}\mathbb{P}_{s}\left(k_{i}^{-2}\left(\hat{y}_{i}-y_{i}\right)y_{i}^{\top}\right)\right)P,
\end{split}
\label{eq:GAME}
\end{equation}
where the map $\textbf{E}$ is given in \eqref{eq:E}. We note that
the $H_{\infty}$ and GAME filters use the same innovation terms,
but differ significantly in their gain update equations. More precisely,
the $H_{\infty}$ filter contains a $\gamma$-related term, whereas
the GAME filter contains geometric curvature correction terms which
enable it to have better transient performance than the MEKF.
\end{rem}
\begin{rem}
As mentioned earlier, the storage function \eqref{eq:StorageFunction_Axang}
can be used to obtain stability and performance guarantees only for
angular errors up to $\pi/2$ radians. A better starting point would
be to assume a storage function which is strictly increasing in the
angular error. More precisely, consider the following chordal metric-based
storage function:
\begin{equation}
V_{\text{ch}}\left(R,\hat{R},t\right)=\frac{\gamma^{2}}{2}\text{tr}\left[P^{-1}\left(t\right)\left(I-\hat{R}^{\top}R\right)\right]\label{eq:StorageFunction_Chordal}
\end{equation}
From \cite[Lemma 2.2.5]{BerkaneThesis}, we observe that
\begin{equation}
V_{\text{ch}}=\frac{1}{2}\gamma^{2}\left(1-\cos\tilde{\theta}\right)e^{\top}\textbf{E}\left(P^{-1}\right)e,\label{eq:StorageFunction_Chordal_Axang}
\end{equation}
where the map $\textbf{E}$ is given in \eqref{eq:E}. We note that
a similar function is used in \cite{zamani2011near} for deriving
a minimum energy filter on $SO(3)$. In future work, we hope to re-visit
the problem of worst-case $H_{\infty}$ estimation using a chordal
metric-based storage function.
\end{rem}

\section{Simulation Results\label{sec:Simulation-Results}}

In this section we simulate the proposed filter \eqref{eq:EHF}, as
well as the GAME filter and the MEKF, against different test cases
and compare their performance. The proposed filter and the minimum
energy filter are both implemented using a full rotation matrix representation
for the state with vector measurements, while the MEKF is implemented
in using the quaternion representation and vector measurements \eqref{eq:MEKF_Estimator}-\eqref{eq:Riccati_MEKF}.

Two different test cases are considered. In Section \ref{subsec:Case-A}
we consider a simulation case whose parameters are motivated by the
characteristics of a typical inertial measurement unit (IMU) used
in small quad-rotors. In Section \ref{subsec:Case-B}, the test case
is characterized by a relatively higher process (\emph{i.e}., gyro)
noise but lower measurement noise.

\subsection{Case A\label{subsec:Case-A}}

\begin{figure}[tbh]
\includegraphics[width=0.9\columnwidth]{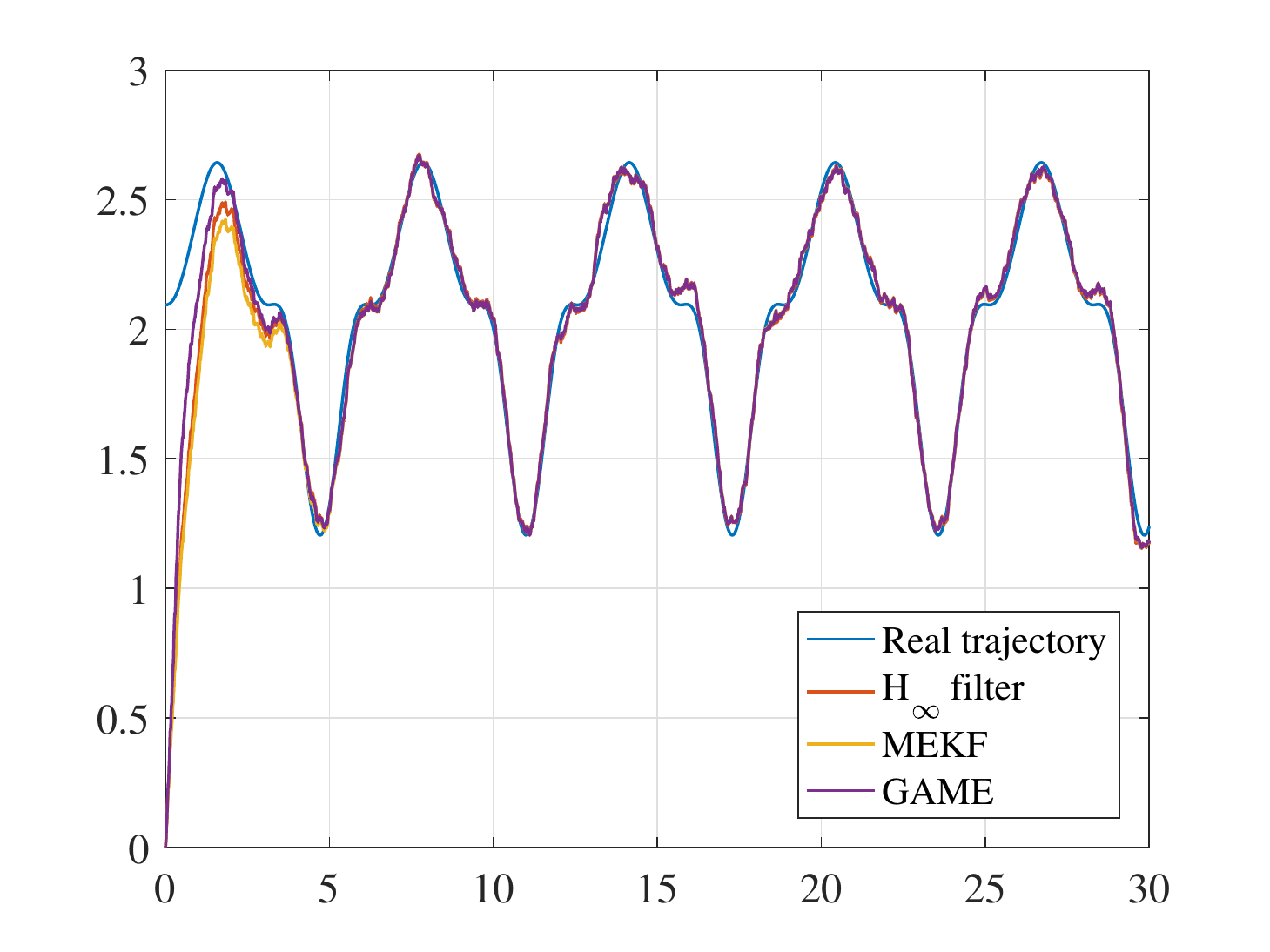}\caption{True and estimated trajectories (Case A)\label{fig:Trajs_Case_A}}
\end{figure}

The true trajectory of the system is generated using
\begin{align*}
\dot{R} & =R\omega_{\text{true}}^{\times}, & R\left(0\right)=R_{0},
\end{align*}
where $R_{0}$ is the rotation matrix corresponding to the Euler angles
$\begin{bmatrix}\pi & -\frac{\pi}{2} & \frac{\pi}{2}\end{bmatrix}^{\top}$,
and $\omega_{\text{true}}$ is the true angular velocity, given by
\begin{align*}
\omega_{\text{true}} & =\begin{bmatrix}\cos3t\\
0.1\sin2t\\
-\cos t
\end{bmatrix}\text{rad/s}.
\end{align*}
The angular velocity signal is then integrated to obtain the true
trajectory of the state on $SO(3)$. This true trajectory, see Figure
\ref{fig:Trajs_Case_A}, is clearly simulating a large movement on
$SO(3)$. The filter, however, only has been provided with the corrupted
angular velocity signal, similar to the case of gyroscopic measurements
of angular rate. The process disturbance corrupting the true angular
velocity is generated by a vector of three independent sources of
zero-mean white noises having standard deviations of $\sqrt{\pi/12}$
rad/s. The measurement disturbance vectors are obtained by generating
two sets of the three independent realizations of zero-mean white
noises having standard deviations of $\sqrt{\pi/12}$ rad. The system
and the filters are both simulated with a time step of 0.01 seconds
(\emph{i.e.}, at 100 Hz), and for a total simulation time of 30 seconds.

To get a better insight into the convergence and tracking performance
of the filters, the system model is initialized with a large initial
attitude, equivalent to the Euler angles $\begin{bmatrix}\pi & -\frac{\pi}{2} & \frac{\pi}{2}\end{bmatrix}^{T}$.
The filters are however assumed to have no prior knowledge of the
initial attitude, and to cope with this uncertainty the initial gain
is set to a large value, \emph{i.e.}, $P_{0}=\frac{1}{2}I$. The MEKF
is tuned with the true parameters of process and measurement disturbance.
Although the proposed $H_{\infty}$ filter and the GAME filter are
deterministic filters, the knowledge of process and measurement noises
can still be used to tune these filters. Hence, we choose $g=\sigma_{\delta}$
and $k_{i}=\sigma_{\epsilon}$ for both of these filters. Furthermore,
for the $H_{\infty}$ filter, $\gamma=0.9$ is found to offer a good
trade-off between stability and speed of convergence. While decreasing
$\gamma$ further means a tighter bound on the $L_{2}$ error gain,
it is to be reminded that a solution of the Riccati equation \eqref{eq:EHF}
may cease to exist for too small values of $\gamma$, see \emph{e.g.},
\cite{simon2006optimal,lewis2007estimation}.
\begin{figure}[tbh]
\includegraphics[width=0.9\columnwidth]{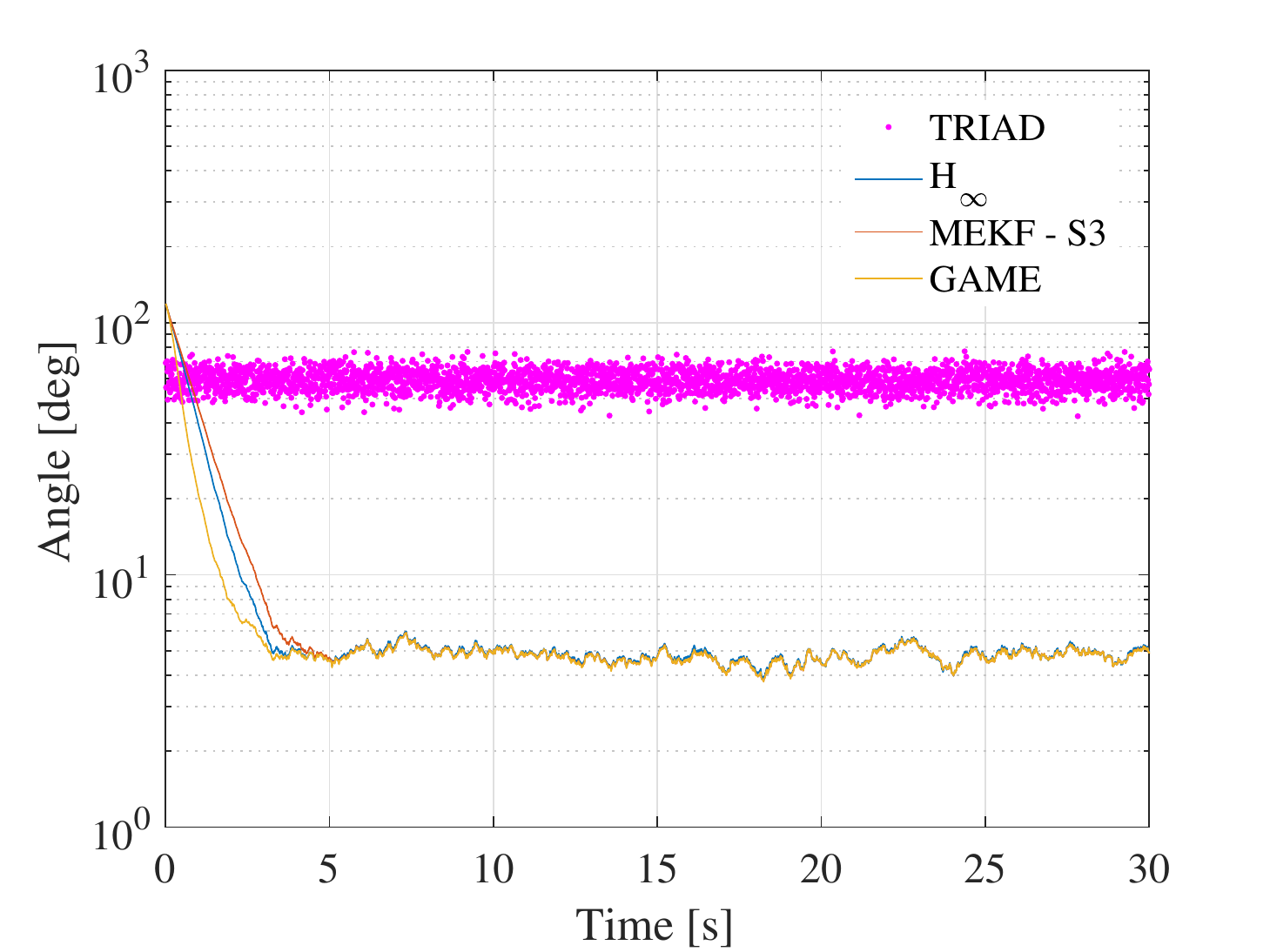}\caption{Measurement and estimation errors - RMS values (Case A)\label{fig:Errors_Case_A}}
\end{figure}
\begin{table}[htbp]
\caption{Estimation errors (deg) for Case A\label{tab:Estimation-errors-RMS_Table}}

\centering{}%
\begin{tabular}{ccc}
\toprule
 & Transient & Steady state\tabularnewline
 & RMS error & RMS error\tabularnewline
\midrule
\midrule
TRIAD & 59.52 & 59.29\tabularnewline
\midrule
MEKF & 27.79 & 4.74\tabularnewline
\midrule
$H_{\infty}$ Filter & 26.24 & 4.79\tabularnewline
\midrule
GAME & 21.68 & 4.73\tabularnewline
\bottomrule
\end{tabular}
\end{table}

The absolute values of the estimation errors are averaged over 50
runs to obtain the plots of estimation error, see Figure \ref{fig:Errors_Case_A},
and to compute the normalized RMS error in Table \ref{tab:Estimation-errors-RMS_Table}.
The RMS error is split in two parts: the transient part during the
first 10 seconds and the steady state part after the transient part.
The results suggest that the GAME filter demonstrates the best performance
during the transients while the proposed filter \eqref{eq:EHF} performs
slightly better than the MEKF. All the filters seem to exhibit similar
performance at the steady-state. We conjecture that its the effect
of geometric curvature terms in the Riccati equation \eqref{eq:GAME},
which makes the GAME filter more agile during the transients.

\subsection{Case B\label{subsec:Case-B}}

Now we consider another case which is characterized by a high process
noise (which signifies a less accurate system model) but a low measurement
noise. The same simulation setup is maintained as in Section \ref{subsec:Case-A},
except that process and measurement disturbances are taken to be $2\cdot\sqrt{\pi/12}$
rad/s and $\frac{1}{2}\cdot\sqrt{\pi/12}$ rad, respectively. As in
the previous section, the filters are tuned with the actual noise
levels, and the initial conditions are assumed to be unknown.
\begin{figure}[tbh]
\includegraphics[width=0.9\columnwidth]{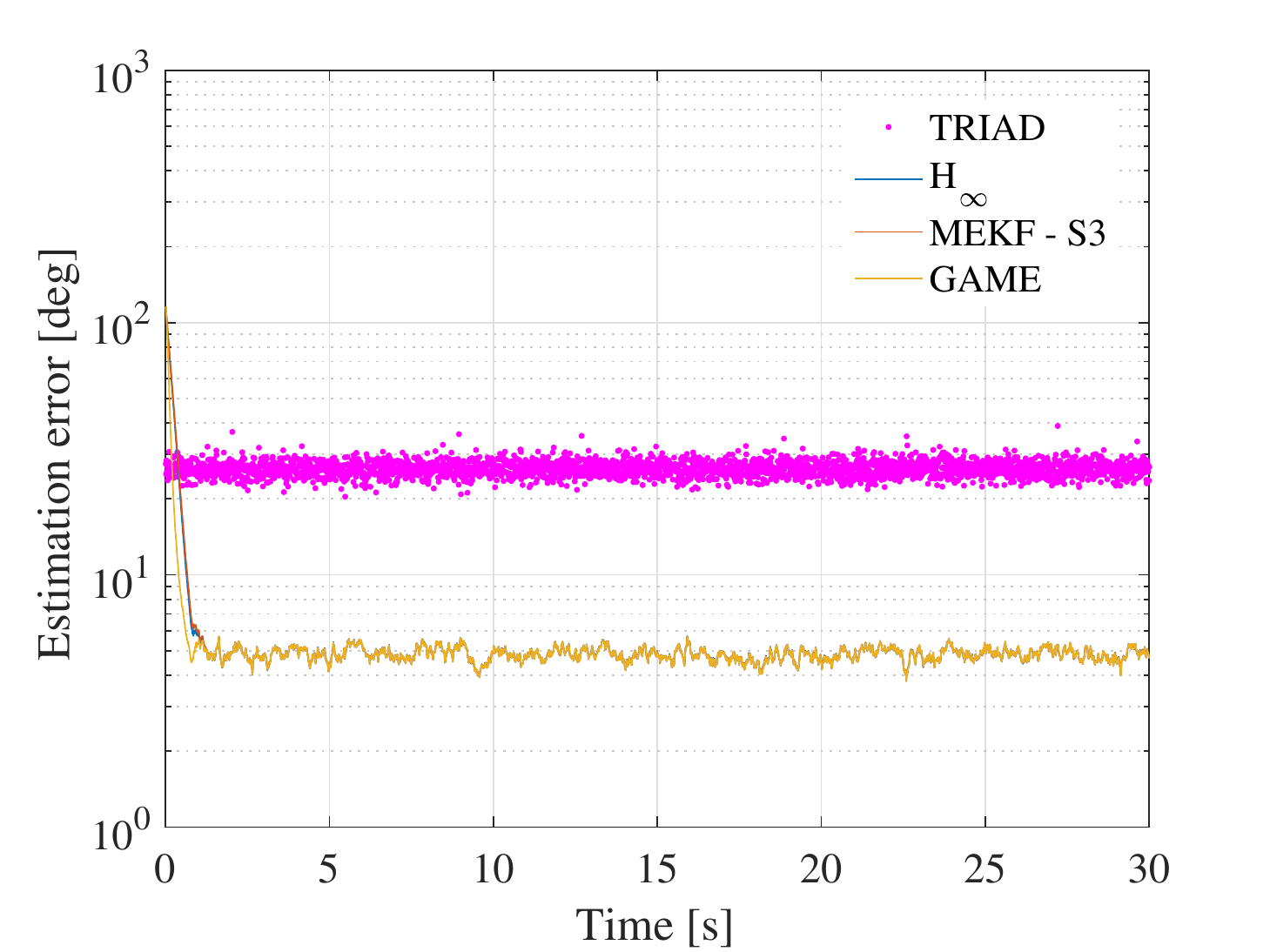}\caption{Measurement and estimation errors - RMS values (Case B)\label{fig:Errors_Case_B}}
\end{figure}
\begin{table}[htbp]
\caption{Estimation errors (deg) for Case B\label{tab:Error_RMS_Case_B}}

\centering{}%
\begin{tabular}{ccc}
\toprule
 & Transient & Steady state\tabularnewline
 & RMS error & RMS error\tabularnewline
\midrule
\midrule
TRIAD & 26.33 & 26.43\tabularnewline
\midrule
MEKF & 14.82 & 4.84\tabularnewline
\midrule
$H_{\infty}$ Filter & 14.63 & 4.85\tabularnewline
\midrule
GAME & 11.85 & 4.84\tabularnewline
\bottomrule
\end{tabular}
\end{table}

The simulation results for this case are reported in Figure \ref{fig:Errors_Case_B}
and Table \ref{tab:Error_RMS_Case_B}. We see that the GAME filter
is still the winner during the transients, while the proposed $H_{\infty}$
filter is just a little better than the MEKF. At the steady-state,
however, all the filters have comparable performance.
\begin{rem}
It is a common practice that the filter gain is initialized with a
sufficiently high numerical value which leads to faster convergence.
It has been observed after several simulation runs that GAME filter,
being a more agressive filter due to the geometric curvature correction
terms, does not require such high numerical initialization. Instead,
such a high gain initialization may even make GAME filter unstable
in some extreme cases.
\end{rem}

\section{Conclusion\label{sec:Conclusion}}

The problem of $H_{\infty}$ filtering on $SO(3)$ using vector directions
has been investigated in this brief. A deterministic nonlinear $H_{\infty}$
filter has been derived which ensures that a dissipation inequality
is satisfied for estimation errors up to $\pi/2$ radians, and that
the energy gain from exogenous disturbances and estimation errors
to a specified performance measure respects a given upper bound. The
proposed filter has been compared with the quaternion MEKF, the quaternion
$H_{\infty}$ filter, and the Geometric Approximate Minimum Energy
(GAME) filter. The comparison suggests that the proposed filter can
be viewed as a MEKF variant which can achieve better transient performance
without significant degradation in steady-state performance. However,
the approach used to obtain the filter has its limitations. In particular,
the storage function considered can give, at best, only local stability
and performance guarantees. In future work, we hope to re-visit the
problem of worst-case $H_{\infty}$ estimation on $SO(3)$ using a
chordal metric-based storage function, as well as to incorporate geometric
curvature correction terms, such as those used in the GAME filter,
in the gain update equation.

\bibliographystyle{IEEEtran}
\bibliography{IEEEabrv,IEEEexample,Library}

\begin{thebibliography}{10}
\providecommand{\url}[1]{#1}
\csname url@samestyle\endcsname
\providecommand{\newblock}{\relax}
\providecommand{\bibinfo}[2]{#2}
\providecommand{\BIBentrySTDinterwordspacing}{\spaceskip=0pt\relax}
\providecommand{\BIBentryALTinterwordstretchfactor}{4}
\providecommand{\BIBentryALTinterwordspacing}{\spaceskip=\fontdimen2\font plus
\BIBentryALTinterwordstretchfactor\fontdimen3\font minus
  \fontdimen4\font\relax}
\providecommand{\BIBforeignlanguage}[2]{{%
\expandafter\ifx\csname l@#1\endcsname\relax
\typeout{** WARNING: IEEEtran.bst: No hyphenation pattern has been}%
\typeout{** loaded for the language `#1'. Using the pattern for}%
\typeout{** the default language instead.}%
\else
\language=\csname l@#1\endcsname
\fi
#2}}
\providecommand{\BIBdecl}{\relax}
\BIBdecl

\bibitem{markley2005nonlinear}
F.~L. Markley, J.~L. Crassidis, and Y.~Cheng, ``Nonlinear attitude filtering
  methods,'' in \emph{AIAA Guidance, Navigation, and Control Conference}, 2005,
  pp. 15--18.

\bibitem{crassidis2007survey}
J.~L. Crassidis, F.~L. Markley, and Y.~Cheng, ``Survey of nonlinear attitude
  estimation methods,'' \emph{Journal of Guidance, Control, and Dynamics},
  vol.~30, no.~1, pp. 12--28, 2007.

\bibitem{van1992sub}
A.~J. van~der Schaft, ``{$L_2$}-gain analysis of nonlinear systems and
  nonlinear state-feedback $\uppercase{H}_{\infty}$ control,'' \emph{IEEE
  Transactions on Automatic Control}, vol.~37, no.~6, pp. 770--784, 1992.

\bibitem{van1993nonlinear}
------, ``Nonlinear state space {$H_\infty$} control theory,'' in \emph{Essays
  on Control}.\hskip 1em plus 0.5em minus 0.4em\relax Springer, 1993, pp.
  153--190.

\bibitem{isidori1994h}
A.~Isidori, ``{$H_\infty$} control via measurement feedback for affine
  nonlinear systems,'' \emph{International Journal of Robust and Nonlinear
  Control}, vol.~4, no.~4, pp. 553--574, 1994.

\bibitem{krener1994necessary}
A.~Krener, ``Necessary and sufficient conditions for nonlinear worst case
  ({$H_\infty$}) control and estimation,'' \emph{Journal of Mathematical
  Systems, Estimation, and Control}, vol.~4, no.~4, pp. 1--25, 1994.

\bibitem{berman1996h}
N.~Berman and U.~Shaked, ``{$H_\infty$} nonlinear filtering,''
  \emph{International Journal of Robust and Nonlinear Control}, vol.~6, no.~4,
  pp. 281--296, 1996.

\bibitem{aguiar2009robust}
A.~P. Aguiar and J.~P. Hespanha, ``Robust filtering for deterministic systems
  with implicit outputs,'' \emph{Systems \& Control letters}, vol.~58, no.~4,
  pp. 263--270, 2009.

\bibitem{markley1993h}
F.~Markley, N.~Berman, and U.~Shaked, ``{$H_\infty$}-type filter for spacecraft
  attitude estimation,'' \emph{Spaceflight Dynamics 1993}, pp. 697--711, 1993.

\bibitem{markley1994deterministic}
------, ``Deterministic {EKF}-like estimator for spacecraft attitude
  estimation,'' in \emph{American Control Conference, 1994}, vol.~1.\hskip 1em
  plus 0.5em minus 0.4em\relax IEEE, pp. 247--251.

\bibitem{cooper2010spacecraft}
L.~Cooper, D.~Choukroun, and N.~Berman, \emph{Spacecraft Attitude Estimation
  via Stochastic {$H_\infty$} Filtering}.\hskip 1em plus 0.5em minus
  0.4em\relax American Institute of Aeronautics and Astronautics, 2010.

\bibitem{choukroun2011spacecraft}
D.~Choukroun, L.~Cooper, and N.~Berman, ``Spacecraft attitude estimation and
  gyro calibration via stochastic {$H_\infty$} filtering,'' in \emph{Advances
  in Aerospace Guidance, Navigation and Control}.\hskip 1em plus 0.5em minus
  0.4em\relax Springer, 2011, pp. 397--415.

\bibitem{chaturvedi2011rigid}
N.~A. Chaturvedi, A.~K. Sanyal, and N.~H. McClamroch, ``Rigid-body attitude
  control,'' \emph{IEEE Control Systems}, vol.~31, no.~3, pp. 30--51, 2011.

\bibitem{mahony2005complementary}
R.~Mahony, T.~Hamel, and J.-M. Pflimlin, ``Complementary filter design on the
  special orthogonal group {SO}(3),'' in \emph{44th IEEE Conference on Decision
  and Control, and European Control Conference. CDC-ECC'05.}\hskip 1em plus
  0.5em minus 0.4em\relax IEEE, 2005, pp. 1477--1484.

\bibitem{mahony2008nonlinear}
------, ``Nonlinear complementary filters on the special orthogonal group,''
  \emph{IEEE Transactions on Automatic Control}, vol.~53, no.~5, pp.
  1203--1218, 2008.

\bibitem{berkane2016deterministic}
S.~Berkane and A.~Tayebi, ``On deterministic attitude observers on the
  \uppercase{S}pecial \uppercase{O}rthogonal group {SO}(3),'' in \emph{55th
  IEEE Conference on Decision and Control (CDC)}.\hskip 1em plus 0.5em minus
  0.4em\relax IEEE, 2016, pp. 1165--1170.

\bibitem{berkane2017design}
------, ``On the design of attitude complementary filters on
  $\uppercase{SO}(3)$,'' \emph{IEEE Transactions on Automatic Control},
  vol.~63, no.~3, pp. 880--887, 2017.

\bibitem{bonnabel2008symmetry}
S.~Bonnabel, P.~Martin, and P.~Rouchon, ``Symmetry-preserving observers,''
  \emph{IEEE Transactions on Automatic Control}, vol.~53, no.~11, pp.
  2514--2526, 2008.

\bibitem{lageman2010gradient}
C.~Lageman, J.~Trumpf, and R.~Mahony, ``Gradient-like observers for invariant
  dynamics on a ${L}$ie group,'' \emph{IEEE Transactions on Automatic Control},
  vol.~55, no.~2, pp. 367--377, 2010.

\bibitem{bonnable2009invariant}
S.~Bonnabel, P.~Martin, and E.~Sala{\"u}n, ``Invariant \uppercase{E}xtended
  \uppercase{K}alman \uppercase{F}ilter: theory and application to a
  velocity-aided attitude estimation problem,'' in \emph{Proceedings of the
  48th IEEE Conference on Decision and Control, 2009.}\hskip 1em plus 0.5em
  minus 0.4em\relax IEEE, 2009, pp. 1297--1304.

\bibitem{zamani2011near}
M.~Zamani, J.~Trumpf, and R.~Mahony, ``Near-optimal deterministic filtering on
  the rotation group,'' \emph{IEEE Transactions on Automatic Control}, vol.~56,
  no.~6, pp. 1411--1414, 2011.

\bibitem{zamani2012second}
------, ``A second order minimum-energy filter on the special orthogonal
  group,'' in \emph{American Control Conference (ACC), 2012}.\hskip 1em plus
  0.5em minus 0.4em\relax IEEE, 2012, pp. 1895--1900.

\bibitem{zamani2013minimum}
------, ``Minimum-energy filtering for attitude estimation,'' \emph{IEEE
  Transactions on Automatic Control}, vol.~58, no.~11, pp. 2917--2921, 2013.

\bibitem{ZamaniThesis}
M.~Zamani, ``Deterministic attitude and pose filtering, an embedded
  $\text{Lie}$ groups approach,'' Ph.D. dissertation, Australian National
  University, Canberra, Australia, 2013.

\bibitem{saccon2016second}
A.~Saccon, J.~Trumpf, R.~Mahony, and A.~P. Aguiar, ``Second-order-optimal
  minimum-energy filters on ${L}$ie groups,'' \emph{IEEE Transactions on
  Automatic Control}, vol.~61, no.~10, pp. 2906--2919, 2016.

\bibitem{zamani2015nonlinear}
M.~Zamani, J.~Trumpf, and R.~Mahony, ``Nonlinear attitude filtering: a
  comparison study,'' \emph{arXiv preprint arXiv:1502.03990}, 2015.

\bibitem{lavoie2019invariant}
M.-A. Lavoie, J.~Arsenault, and J.~R. Forbes, ``An \uppercase{I}nvariant
  \uppercase{E}xtended $\uppercase{H}_{\infty}$ \uppercase{F}ilter,'' in
  \emph{2019 IEEE 58th Conference on Decision and Control (CDC)}.\hskip 1em
  plus 0.5em minus 0.4em\relax IEEE, 2019, pp. 7905--7910.

\bibitem{haydarCDC17}
M.~F. Haydar and M.~Lovera, ``$\uppercase{H}_{\infty}$ filtering on the unit
  circle,'' in \emph{2017 IEEE 56th Annual Conference on Decision and Control
  (CDC)}.\hskip 1em plus 0.5em minus 0.4em\relax IEEE, 2017, pp. 2422--2427.

\bibitem{markley2004multiplicative}
F.~L. Markley, ``Multiplicative vs. additive filtering for spacecraft attitude
  determination,'' \emph{Dynamics and Control of Systems and Structures in
  Space}, pp. 467--474, 2004.

\bibitem{gui2018quaternion}
H.~Gui and A.~H. de~Ruiter, ``Quaternion invariant extended $\text{K}$alman
  filtering for spacecraft attitude estimation,'' \emph{Journal of Guidance,
  Control, and Dynamics}, vol.~41, no.~4, pp. 863--878, 2018.

\bibitem{izadi2015comparison}
M.~Izadi, E.~Samiei, A.~K. Sanyal, and V.~Kumar, ``Comparison of an attitude
  estimator based on the $\text{L}$agrange-d'$\text{A}$lembert principle with
  some state-of-the-art filters,'' in \emph{2015 IEEE International Conference
  on Robotics and Automation (ICRA)}.\hskip 1em plus 0.5em minus 0.4em\relax
  IEEE, 2015, pp. 2848--2853.

\bibitem{BerkaneThesis}
S.~Berkane, ``Hybrid attitude control and estimation on {$SO(3)$},'' Ph.D.
  dissertation, University of Western Ontario, London, Canada, 2017.

\bibitem{simon2006optimal}
D.~Simon, \emph{Optimal state estimation: Kalman, {$H_\infty$}, and nonlinear
  approaches}.\hskip 1em plus 0.5em minus 0.4em\relax John Wiley \& Sons, 2006.

\bibitem{lewis2007estimation}
F.~L. Lewis, L.~Xie, and D.~Popa, \emph{Optimal and robust estimation: with an
  introduction to stochastic control theory}.\hskip 1em plus 0.5em minus
  0.4em\relax CRC press, 2007, vol.~29.

\end{thebibliography}

\end{document}